\documentclass[preprint, 5p, times, twocolumn]{elsarticle}

\usepackage{amssymb}
\usepackage{amsmath}
\usepackage{listings}
\usepackage{xcolor}
\usepackage{minted}
\usepackage[geometry]{ifsym}
\usepackage[T1]{fontenc} 
\usepackage{algorithm}
\usepackage{algpseudocode}
\usepackage{tikz}
\usetikzlibrary{shapes.geometric, arrows}
\usepackage{url}
\usepackage{forest}
\usepackage{hyperref}
\usepackage{inconsolata} 

\tikzstyle{startstop} = [rectangle, rounded corners, 
minimum width=3mm, 
minimum height=1mm,
text centered, 
draw=black, 
fill=white]
\tikzstyle{arrow} = [thick,->,>=stealth]

\lstdefinestyle{lammps}{
  escapeinside={(*@}{@*)}, 
  backgroundcolor=\color{gray!5},
  basicstyle=\ttfamily\small,
  numbers=left,
  numberstyle=\tiny\color{gray},
  frame=single,
  breaklines=true,
  postbreak=\mbox{\textcolor{gray}{$\hookrightarrow$}\space},
  keywordstyle=\color{blue},
  commentstyle=\color{gray!70}\itshape,
  showstringspaces=false
}

\newcounter{bla}

\journal{Computer Physics Communications}

\begin{document}

\begin{frontmatter}



\title{NAVIS: A LAMMPS-Python framework for efficient computation of nanochannel velocity and thermal interfacial slip}


\author[a]{Sleeba Varghese\corref{author}}
\author[b]{Sobin Alosious}
\author[c]{J. S. Hansen}
\author[d]{B. D. Todd}

\cortext[author] {Corresponding author.\\\textit{E-mail address:} sleebaslv@gmail.com}
\address[a]{Univ Lyon, INSA-Lyon, CETHIL CNRS-UMR5008, F-69621 Villeurbanne, France}
\address[b]{Aerospace and Mechanical Engineering Department, The University of Notre Dame, Notre Dame, IN 46556, USA}
\address[c]{``Glass and Time", IMFUFA, Department of Science and Environment, Roskilde University, Roskilde 4000, Denmark}
\address[d]{Department of Mathematics, School of Science, Computing and Emerging Technologies, Swinburne University of Technology, John Street, Hawthorn, Victoria 3122, Australia}

\begin{abstract}
We present NAVIS (\textbf{NA}nochannel \textbf{V}elocity and thermal \textbf{I}nterfacial \textbf{S}lip), a LAMMPS-Python scripted toolkit for computing the Navier (hydrodynamic) friction coefficient and Kapitza (thermal) resistance at arbitrary solid-fluid interfaces. 
NAVIS is based on equilibrium molecular dynamics (EMD) methods for calculating the linear response friction and thermal resistance at the interface, as well as the corresponding velocity and temperature slips.
The methodology is based on our previous studies (Hansen, {\it{et al.}}, Phys. Rev. E {\bf{84}}, 016313 (2011); Varghese {\it{et al.}}, J. Chem. Phys. {\bf{154}}, 184707 (2021); Alosious, {\it{et al.}}, J.
Chem. Phys. {\bf{151}}, 194502 (2019); Alosious, {\it{et al.}}, Langmuir {\bf{37}}, 2355-2361
(2021)), and in this work we provide a pedagogical framework for the implementation of this toolkit on two systems: (\romannumeral 1) a water-graphene system (for hydrodynamic slip) and (\romannumeral 2) a water-CNT system (for thermal slip).
We provide detailed instructions for performing the EMD simulations using the LAMMPS package and processing the simulation outputs using Python modules to obtain the desired quantities of interest. We expect the toolkit to be useful for computational researchers studying interfacial friction and thermal transport, key factors for efficient and practical applications of nanofluidic systems.
\end{abstract}


\end{frontmatter}



{\bf PROGRAM SUMMARY}

\begin{small}
\noindent
{\em Program Title:} NAVIS  \\
{\em CPC Library link to program files:} {\color{red}(to be added by Technical Editor)}  \\
{\em Developer's repository link:} \url{https://github.com/sleebaslv/NAVIS} \\
{\em Licensing provisions:} GNU General Public License v3.0  \\
{\em Programming language:} Python 3    \\
{\em Nature of problem:} Measure the interface-intrinsic hydrodynamic and thermal resistance at the solid-fluid boundary using equilibrium molecular dynamics (EMD) simulations.\\
{\em Solution method:} Creating a Python-based code to compute hydrodynamic and thermal interfacial friction coefficients from data obtained via EMD simulations conducted using LAMMPS.
\end{small}

\section{Introduction}
\label{}
The study of friction and heat transfer at the solid-liquid interface has in recent years become a focus of intensified research. At its heart lays the prospect of improving lubrication technologies to minimise energy losses due to friction (tribology) whilst also improving heat exchange efficiency - either to ensure maximum heat exchange for cooling purposes or the opposite if insulation is required. Driving this study is climate change and the urgency to reduce greenhouse gas emissions, whilst simultaneously improving energy efficiency in mechanical devices and batteries. It is not widely known that 23$\%$ of global energy consumption is due to tribological contacts and that 20$\%$ of this is due to the energy required to overcome friction, including the friction generated in electric vehicles \cite{holmberg2019}. Clearly then, increased knowledge of friction and thermal energy transfer at the crucial solid-liquid interface is highly sought after and many studies involving experiment, theory and computer simulation are devoted to this~\cite{neto2005boundary, bocquet2010nanofluidics, ajdari2006giant, sam2021fast,pollack1969kapitza,chen2022interfacial}.

From a conceptual perspective, the friction that exists between the solid wall and fluid can be understood as the ability/inability of the interfacial fluid to experience ``slip'' along the solid surface. Slip is defined as the departure of a corresponding observable (velocity for hydrodynamic slip and temperature for thermal slip) from the no-slip boundary condition, and its effect becomes significant when the confinement width reduces to nanometric scales.
The hydrodynamic and thermal slips are often characterized in terms of interfacial transport coefficients, namely  Navier friction coefficient (for hydrodynamic friction) and Kapitza resistance (for thermal friction), respectively. Hence, efficient design of nanoscale systems requires theoretical/computational frameworks that can provide precise estimates of these transport coefficients, consistent with the nature of the solid-fluid interface under consideration.

Various computational approaches are available to calculate interfacial transport properties such as friction and thermal resistance. Broadly, these methods can be classified into equilibrium molecular dynamics (EMD) approaches, which exploit linear-response theory and yield properties that are directly comparable to experimental observables, and nonequilibrium molecular dynamics (NEMD) approaches, which impose external driving fields to probe the system’s response. While NEMD techniques can access nonlinear regimes, it is unfeasible to generate statistically reliable flow fields at experimentally relevant external field strengths and consequently extrapolation techniques must be employed to obtain friction coefficient values at experimentally achievable field strengths~\cite{todd-daivis2017}. This drawback of NEMD techniques becomes particularly noticable for high-slip systems, as demonstrated by Kannam {\it et al.}~\cite{kannam2012, kannam2013}, making linear-response theory based techniques (EMD) the preferred choice. However recent developments, such as the transient-time correlation function (TTCF) method~\cite{bernardi2012, bernardi2016, maffioli2022, maffioli2024ttcf4lammps}, now enable direct NEMD simulations at physically meaningful rates of strain and can be implemented in conjunction with the LAMMPS package~\cite{maffioli2024ttcf4lammps}. Given the generality and accuracy of the linear-response theory under normal laboratory conditions, in this work we primarily focus on the computation of friction and thermal resistance using EMD simulations.

To our knowledge, the first linear response model to predict the friction coefficient at the solid-fluid interface was that of Bocquet and Barrat \cite{bocquet1994}. Their method involves a Green-Kubo-like integration of a time-autocorrelation function of tangential interfacial forces, performed at equilibrium. Bocquet and Barrat do warn users that such a formulation for confined systems must be treated with caution, noting that: ``... the evaluation of these formulas leads in finite systems to vanishing transport coefficients'' and ``To obtain a nonvanishing transport coefficient... the thermodynamic limit must be taken.'' In short, if one does not truncate the correlation function at some point in delay-time, it diverges and the corresponding integral vanishes, leading to an under-estimate for the friction coefficient and hence slip lengths that are higher than they should be. The source of this divergence is the fact that Green-Kubo methods are intrinsically applicable to systems that are at zero-wavevector, i.e. unconfined ``bulk'' systems, or as Bocquet and Barrat state are in ``the thermodynamic limit''. This divergence is clearly demonstrated in reference \cite{varghese2021} and discussions can be found in references ~\cite{todd-daivis2017, espanol1993force, evans-morriss1990, petravic2007}. For MD simulations this becomes a computational bottleneck. Several authors have proposed practical strategies to address this issue~\cite{brey1982computer, bocquet2013green, huang2014, oga2019, espanol2019, nakano2019, nakano2020, oga2023}, and the reader is referred to these works for more information on this methodology. The theoretical foundations of this non-zero wavevector divergence can be found in the books by Hansen and McDonald \cite{hansen-mcdonald2013}, Evans and Morriss \cite{evans-morriss1990} and Todd and Daivis \cite{todd-daivis2017}. Petravic and Harrowell \cite{petravic2007, petravic2008} also pointed out that by integrating over the entire volume of the confined fluid, the friction coefficient determined includes the viscous resistance of the fluid itself, and so is not an intrinsic friction coefficient. 

In an attempt to remove altogether the controversy of applying Green-Kubo-like methods to confined (non-zero wavevector) systems, Hansen {\it{et al.}} \cite{hansen2011} proposed an alternative linear response (equilibrium MD) method that did not involve an integration over the volume of the fluid. It has the additional advantage of being able to calculate the solid-fluid friction coefficients for a fluid confined by two different solid walls during a single simulation, i.e. obtaining two unique intrinsic friction coefficients reflective of the two different confining surfaces. In contrast, the Bocquet-Barrat method can only obtain an effective friction coefficient for such a system, in which it is a hybrid of both interfaces. Recently, Varghese {\it{et al.}} \cite{varghese2021} proposed an improvement to the method of Hansen {\it{et al.}} that increased its statistical accuracy, and it is this latter implementation we reproduce and implement in this work.  

In an entirely analogous fashion, the calculation of the weak-field (linear response) Kapitza resistance can be formulated by either a Green-Kubo-like correlation function integral proposed by Barrat and Chiaruttini \cite{barrat2003} or by the non-integral time-correlation approach proposed by Alosious {\it{et al.}} \cite{alosious2019, alosious2021}. As with the case of the friction coefficient, the former method involves an integral which {\it{diverges}} at long times, whereas the latter does not, as clearly shown in reference \cite{alosious2021}. In this work, we implement the calculation proposed by Alosious {\it{et al.}} \cite{alosious2021} because of its stability and statistical accuracy.

Although the \emph{statistical mechanics} framework forms the foundation of any EMD approach for computing thermal or velocity slip, an equally important aspect is the effective implementation of this theoretical framework into a practical computational methodology.
A clear and pedagogical implementation of this workflow (Theory $\to$ Code) can help avoid the repeated ``reinvention of the wheel'' when developing new computational tools from established theoretical principles. Such efforts have gained considerable traction in recent years due to the growing use of molecular simulations to probe the underlying physics of complex systems~\cite{maffioli2024ttcf4lammps, gravelle2025set, fish2026libmobility}
Hence, the motive of the present work is to provide a useful computational toolkit, implemented in Python \cite{python} and interfaced with the popular LAMMPS software package \cite{thompson2022}, that can readily compute the Navier friction coefficient and Kapitza resistance for arbitrary solid-fluid interfaces. 
The following section briefly discusses the theoretical framework behind the hydrodynamic and thermal friction computing methodologies. The subsequent sections provide detailed descriptions of the simulation and post-processing procedures used to compute the Navier friction coefficient and Kapitza resistance, followed by a results and discussion section, where we present some important results obtained using the proposed methodologies.

\section{Theory}
\label{sec:theory}
In this section we briefly outline the mathematical formulation to compute the required quantities. Full details of the theory and validation are found in the original papers on slip \cite{hansen2011, varghese2021} and Kapitza resistance \cite{alosious2019, alosious2020, alosious2021, alosious2022}, and the reader is encouraged to refer to them for more details.

The basic idea proposed for the calculation of the Navier friction coefficient and the Kaptiza resistance is that they are both {\it{local}} interfacial properties that are dominated by interactions between molecules in the immediate fluid layer adjacent to the fluid and the corresponding wall atoms. Time-correlation functions are formed between relevant quantities based upon the defining constitutive equation, equilibrium molecular dynamics simulations on the system of interest are performed, and then the Navier friction coefficient or Kapitza resistance is extracted from the relevant time-correlation data. The fundamental expressions are summarised below.

\subsection{Navier Friction coefficient}
\label{sec:hyd_fric_coeff}
\begin{figure}
	\centering	
	\includegraphics[width=1.0\linewidth]{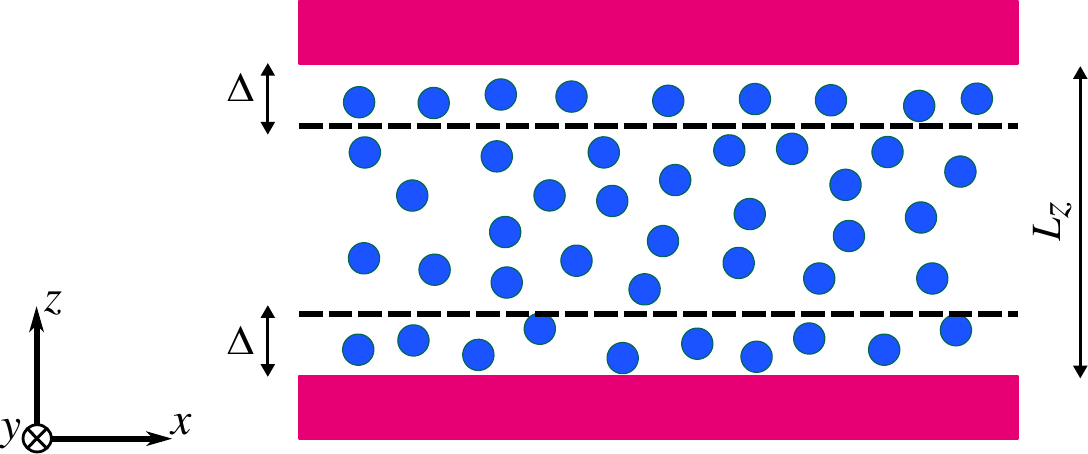}
	\caption{A 2-D schematic diagram of the fluid molecules (depicted as solid circles) confined by planar walls. The actual system is three-dimensional and periodic in the $x$ and $y$ directions. $\Delta$ is the slab width.}
	\label{fig1}
\end{figure}

Let a fluid system be confined between solid walls separated some distance $L_z$ in the $z$-direction, as shown in Figure \ref{fig1}. We assume that the zero-frequency friction coefficient ($\xi_0$) is defined in the usual manner, namely $\sigma_{xy} = \xi_0 u_s$, where $\sigma_{xy}$ and $u_s$ are, respectively, the shear stress and slip velocity measured at the solid-fluid interface. If we assume that any (linear response) flow is in the $x$-direction and the system is periodic in both the $x$ and $y$ directions, then we can define $u_x\left(t\right)$ to be the $x$-component velocity of a slab of fluid of width $\Delta$ in the $z$-direction immediately adjacent to the wall and $F_x\left(t\right)$ to be the $x$-component of force between all fluid atoms in the slab and all wall atoms. If we define the time-correlation functions 
\begin{equation}
\label{eq:corr}
    C_{u_x F_x} \left(t\right) \equiv \left<u_x\left(0\right) F_x\left(t\right)\right>\ \textrm{and}\ 
    C_{u_x u_x} \left(t\right) \equiv \left<u_x\left(0\right) u_x\left(t\right)\right>
\end{equation}
and we denote the Laplace transform of an arbitrary function of time $f\left(t\right)$ as 
\begin{equation}
\label{eq:laplace}
L\left[f\left(t\right)\right] = \int_0^\infty f\left(t\right) e^{-st} dt \equiv \tilde f\left(s\right),    
\end{equation}
then $\xi_0$ is obtained as 
\begin{equation}
\xi_0 = \frac{B_1}{A\lambda_1} \label{friction1}
\end{equation}
where $B_1$ and $\lambda_1$ are parameters of a one-term Maxwellian memory function descriptive of the friction coefficient, defined as
\begin{equation}
    \zeta\left(t\right) = B_1 e^{-\lambda_1 t} \label{friction_kernel}
\end{equation}
and $A$ is the surface area of the solid-liquid interface. $B_1$ and $\lambda_1$ are obtained by Laplace transforming the time-correlation functions obtained from molecular dynamics simulations, as described by Hansen {\it{et al.}} in their original paper \cite{hansen2011}. This leads to the following expression in Laplace-space:
\begin{equation}
    \tilde{C}_{u_x F_x}\left(s\right) = -\frac{B_1}{s+\lambda_1}\tilde{C}_{u_x u_x}\left(s\right). \label{reg1}
\end{equation}

Performing a regression fit to Eq (\ref{reg1}) leads to robust and reliable values of the friction coefficient, from which the slip velocity and slip length may also be accurately predicted. While this is sufficient, Varghese {\it{et al..}} \cite{varghese2021} later demonstrated that even higher statistical accuracy could be obtained by numerically integrating along the Laplacian domain to compute $B_1$ and $\lambda_1$. By integrating Eq (\ref{reg1}) between arbitrary points $s_1$ and $s_2$ and also $s_3$ and $s_4$ in Laplace-space the following two sets of equations are obtained:
\begin{align}
    \frac{1}{B_1}\left[\frac{s_2^2 - s_1^2}{2} + \lambda_1\left(s_2-s_1\right)\right] &= \Lambda_1 \label{simul1}\\
    \frac{1}{B_1}\left[\frac{s_4^2 - s_3^2}{2} + \lambda_1\left(s_4-s_3\right)\right] &=  \Lambda_2 \label{simul2}
\end{align}
where $\Lambda_i \equiv -\int_{s_i}^{s_j}\frac{\tilde{C}_{u_x u_x}\left(s\right)}{\tilde{C}_{u_x F_x}\left(s\right)} ds$ and $j>i$. Defining $C \equiv \Lambda_1/\Lambda_2$ and dividing Eq (\ref{simul1}) by Eq (\ref{simul2}) gives the expression for $\lambda_1$ as
\begin{equation}
    \lambda_1 = \frac{\left(s_2^2 - s_1^2 - C\left(s_4^2 - s_3^2\right)\right)}{2\left(C\left(s_4 - s_3\right) - s_2 + s_1\right)}. \label{lambda1}
\end{equation}
$B_1$ is then easily obtained by substituting $\lambda_1$ into either Eq ({\ref{simul1}}) or Eq ({\ref{simul2}}). It is this latter technique that we use to obtain the friction coefficient $\xi_0$ via Eq.~\eqref{friction1}. 

\subsection{Kaptiza resistance}
Let us assume the same system geometry as above, only there is no assumed momentum flow and the only flow of interest is that of heat. The Kapitza resistance, $R_k$, is defined as $R_k \equiv \Delta T/J_q = 1/G_k$, where $\Delta T$ is the difference between the temperature of the solid interface and a molecularly thin liquid layer immediately adjacent to it, $J_q$ is the interfacial heat flux normal to the surface and $G_k$ is the interfacial thermal conductance. In an analagous manner to the friction coefficient derivation, we find that \cite{alosious2021}
\begin{eqnarray}
    \tilde{C}_{TJ_q}\left(s\right) = \tilde{G}_k\left(s\right) \tilde{C}_{TT}\left(s\right) \label{G_k}
\end{eqnarray}
where $C_{TJ_q}\left(t\right) \equiv \left<T\left(0\right) J_q\left(t\right)\right>$ and $C_{TT}\left(t\right) \equiv \left<T\left(0\right) T\left(t\right)\right>$. Note here that although we follow the specific form of the derivation presented in reference \cite{alosious2021}, an alternative derivation making use of the interfacial heat flux autocorrelation function was the original version of this methodology \cite{alosious2019}. It was later found that using the temperature autocorrelation function yielded statistically superior values of Kapitza resistance and temperature slip predictions, so we only present that version even though both implementations can be used to predict $R_k$ and temperature slip. Expressing the instantaneous $G_k$ as a one-term Maxwellian, $G_k\left(t\right) = k_1 e^{-\mu_1t}$, we find from Eq (\ref{G_k})
\begin{eqnarray}
    \tilde{C}_{TJ_q}\left(s\right) = \frac{k_1}{s+\mu_1} \tilde{C}_{TT}\left(s\right). \label{G_k2}
\end{eqnarray}
$k_1$ and $\mu_1$ are obtained by regression fits to the equilibrium molecular dynamics time-correlation function data, giving us
\begin{eqnarray}
    R_k = \frac{1}{G_k} = \frac{\mu_1}{k_1}. \label{R_k}
\end{eqnarray}
The reader is referred to Ref~\citenum{alosious2019} for further details, including how to compute $J_q$ at the interface, and to Ref~\citenum{alosious2021}  to know more about the implementation of this methodology to cylindrical wall geometries. It is interesting that the Kapitza resistance computed by this method is statistically superior to the friction coefficient computed by the analogous method first proposed by Hansen {\it{et al.}} \cite{hansen2011}. For this reason, we have found no need to extend this method to a form similar to that of Varghese {\it{et al.}}'s \cite{varghese2021} method to enhance the statistical accuracy of the friction coefficient calculation. Computation of the so-called temperature slip will also be demonstrated later.
Unless explicitly stated otherwise, from hereon the term ``slip'' is to be understood as denoting either hydrodynamic (velocity) or thermal (temperature) slip.

\section{Computational Methodology}
\begin{figure}
\centering
\begin{tikzpicture}[node distance=1.5cm]
\node (start) [startstop] {Computational Model};
\node (step1) [startstop, below of=start, align=center] {Simulations \\ (Equilibrium Molecular Dynamics)};
\node (step2) [startstop, below of=step1] {Post-processing};
\node (step3) [startstop, below of=step2] {Navier friction coefficient/Kapitza resistance};

\draw [arrow] (start) -- node[anchor=west] {} (step1);
\draw [arrow] (step1) -- node[anchor=west] {} (step2);
\draw [arrow] (step2) -- node[anchor=west] {} (step3);
\end{tikzpicture}
\caption{Flowchart outlining the steps involved in the computation of the Navier friction coefficient or Kapitza resistance.}
\label{fig:flowchart}
\end{figure}
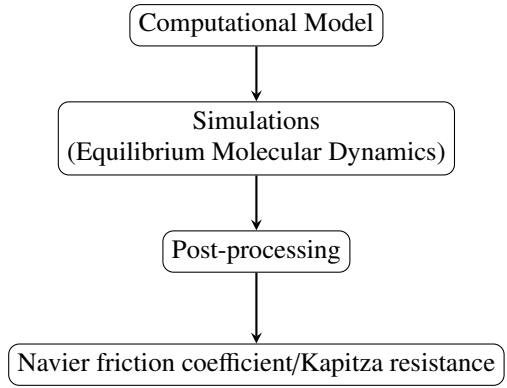

In this section, we explain the simulation and post-processing methodologies used to compute the interfacial slip in a nanoconfined system. Fig~\ref{fig:flowchart} shows the essential steps involved in our NAVIS framework, and in the subsequent subsections we explain each of these steps in detail. The codes and scripts corresponding to the following sections are publicly accessible via GitHub, and the repository structure is described in \ref{sec:App_rep}. 

\subsection{Model}
The effect of interfacial slip becomes a contributing factor to the fluid flow or heat-exchange rate only when the confinement width reduces to nanometeric scales~\cite{kannam2013, bocquet2010nanofluidics}. In our previous works, we used the NAVIS methodology to compute the interfacial slip for different fluids confined in planar  nanochannels \cite{kannam2011, varghese2021, varghese2024existence, alosious2020} and cylindrical nanopores \cite{kannam2012interfacial, kannam2013, alosious2021}. The results presented in this work are taken from either of these previous studies, and the reader is referred to the above mentioned articles for additional details on the system setup for the specific geometry of choice.
Once the computational model has been built, the next step is to
perform equilibrium simulations of the model to measure the 
fluctuations of the relevant properties.

\subsection{Simulations}
For equilibrium simulations, we use molecular dynamics (MD) as our simulation protocol. MD simulations involve numerically solving the Newtonian equations of motion for each particle in the system using appropriate numerical schemes~\cite{frenkel1957understanding, allen2017computer} and for the purpose of our studies we used the Large-scale Atomic/Molecular Massively Parallel Simulator~\cite{thompson2022lammps} (LAMMPS) package for performing the simulations.
The systems are maintained at thermal equilibrium using the thermostatting schemes available in LAMMPS. In our simulations, thermal equilibration is achieved by thermostatting the confining walls rather than the fluid itself, a strategy that has been shown to be the appropriate methodology when working with nanoconfined systems~\cite{bernardi2010thermostating, de2014new, sam2017water}.
Algorithm~\ref{alg:LAMMPS} shows the relevant steps involved in an EMD simulation in LAMMPS to extract the relevant quantities for computing hydrodynamic or thermal slip.

\begin{algorithm}
\caption{LAMMPS simulation procedure for interfacial slip}
\label{alg:LAMMPS}
\begin{algorithmic}[1]
\State Initialize the system.
\State Run $N_{eq}$ timesteps for equilibration.
\State Define the slab region.
\State Define the slab as a dynamic group with the atoms being updated at every timestep.
\If{slip = hydrodynamic slip}
\For{each step $t$ from 1 to $N$}
\State Compute the force acting on the slab due to the wall. 
\State Compute the center of mass velocity of the slab.
\State Save the force component and 
    the center of mass velocity.
\EndFor
\EndIf
\If{slip = thermal slip}
\For{each step $t$ from 1 to $N$}
\State Compute the temperature difference between the solid and fluid slab.
\State Compute the heat flux between the solid and fluid slab.
\State Save the heat flux and temperature difference.
\EndFor
\EndIf

\end{algorithmic}
\end{algorithm}

As mentioned in Sec~\ref{sec:theory}, one of the hallmarks of our slip methodology is that we restrict our zone of interest to a few molecular diameters away from the wall. This avoids the possible interference of the bulk fluid contributions and hence makes the slip estimated using this methodology an intrinsic parameter of the solid-liquid interface. To be consistent with the theoretical framework proposed by Hansen {\it{et al.}}~\cite{hansen2011}, the simulations must  ensure that the relevant quantities are only averaged over the interfacial region of particles. This objective is achieved using the \textit{dynamic group} feature available in LAMMPS.
The code snippet shown in Listing~\ref{code:lammps_navier} provides the syntax for implementing this feature to compute the Navier friction coefficient. Similarly, Listing~\ref{code:lammps_kapitza} shows the code snippet for the Kapitza resistance.
We also mention here that we use the terminology ``slab'' to define the interfacial region and below we provide the relevant quantities that are measured from EMD simulations,\\
\textit{Hydrodynamic slip}:
(\romannumeral 1) Fluctuations of the tangential component of the force acting on the slab due to the wall [compute group/group].
(\romannumeral 2) The tangential component of the relative velocity between the wall and the slab (equivalently, the center-of-mass velocity of the fluid particles within the interfacial region) [variable vcm].\\
\textit{Thermal slip}: 
(\romannumeral 1) Fluctuations of the heat flux vector component normal to the wall.
(\romannumeral 2) The temperature difference between the slab of water immediately adjacent to the wall.\\

\begin{lstlisting}[style=lammps, caption={LAMMPS syntax for creating dynamic groups. Text in $italics$ must be specified by the user accordingly as per the LAMMPS documentation.}, label={code:lammps_navier}]
region      (*@\it{slab\_region\_id}@*) block INF INF INF INF  (*@\it{slab\_zmin}@*)  (*@\it{slab\_zmax}@*) units box 
group       (*@\it{slab\_group\_id}@*) dynamic (*@\it{fluid\_region\_id}@*)  region (*@\it{slab\_region\_id}@*) every (*@\it{update\_freq}@*)
variable    (*@\it{variable\_id}@*) equal vcm((*@\it{slab\_group\_id}@*),x) 
compute     (*@\it{compute\_id}@*)  (*@\it{slab\_group\_id}@*) group/group (*@\it{wall\_group\_id}@*) 
fix     (*@\it{fix\_id}@*) all ave/time 1 1 1 c_(*@\it{compute\_id}@*)[1]  v_(*@\it{variable\_id}@*) (*@\it{output\_filename}@*) mode scalar 
\end{lstlisting}

\begin{lstlisting}[style=lammps, caption={LAMMPS syntax for creating dynamic groups, heat flux, and temperature difference calculation. Text in $italics$ must be specified by the user accordingly as per the LAMMPS documentation.}, label={code:lammps_kapitza}]
region      (*@\it{slab\_region\_id}@*) block INF INF INF INF  (*@\it{slab\_zmin}@*)  (*@\it{slab\_zmax}@*) units box 
group       (*@\it{slab\_group\_id}@*) dynamic (*@\it{fluid\_group\_id}@*)  region (*@\it{slab\_region\_id}@*) every (*@\it{update\_freq}@*)
compute     (*@\it{compute\_id1}@*)  (*@\it{slab\_group\_id}@*) property/atom vz
compute     (*@\it{compute\_id2}@*)  (*@\it{slab\_group\_id}@*) force/tally (*@\it{wall\_group\_id}@*)
variable    (*@\it{variable\_id1}@*) atom c_(*@\it{compute\_id1}@*)*c_(*@\it{compute\_id2}@*)[3]
compute     (*@\it{compute\_id3}@*)  (*@\it{slab\_group\_id}@*) reduce sum v_(*@\it{variable\_id1}@*)
compute     (*@\it{compute\_id4}@*)  (*@\it{wall\_group\_id}@*) temp
compute     (*@\it{compute\_id5}@*)  (*@\it{slab\_group\_id}@*) temp
variable    (*@\it{variable\_id2}@*) equal c_(*@\it{compute\_id4}@*)-c_(*@\it{compute\_id5}@*)
fix         (*@\it{fix\_id}@*) all ave/time 1 1 1 c_(*@\it{compute\_id3}@*)  v_(*@\it{variable\_id2}@*) (*@\it{output\_filename}@*) mode scalar 
\end{lstlisting}

\subsection{Post-processing}
After obtaining adequate data from the EMD simulations, as described in the previous section, we advance to the post-processing stage. In this stage, the data obtained from the EMD simulations are used to calculate the interfacial friction coefficient and/or Kapitza resistance. The workflow to extract the corresponding parameters are:
\begin{enumerate}
    \item Perform the time correlation for the simulation data sets.
    \begin{enumerate}
        \item Navier friction coefficient: Perform auto-correlation of the center-of-mass slab velocity dataset and cross-correlation between the force component and slab velocity datasets (Algorithm~\ref{alg:corr}). The function defined in Algorithm~\ref{alg:corr} takes as input the total time series data for the quantities whose correlations are to be computed. The total time series is divided into \textit{no\_of\_sets} independent blocks, and the final correlation data is the average over these independent blocks. 
        The output of this function consists of the raw time correlation data and the normalized correlation data.
        At present the numerical computation of the auto(cross)-correlation functions (Algorithm~\ref{alg:corr}) has a computational complexity of $\mathcal{O}(N^2)$. This step can be further accelerated to $\mathcal{O}(N\ \mathrm{ln}\ N)$ by evaluating the correlations via the Wiener-Khintchin theorem together with the Fast Fourier Transform~\cite{press2007numerical}.
        \item Kapitza resistance: Perform autocorrelation between the temperature difference data set and cross-correlation between the heat flux-temperature difference data set. The correlation analysis follows the same procedure and data-blocking approach outlined for the Navier friction coefficient in Algorithm~\ref{alg:corr}.
    \end{enumerate}
    \item Transform the correlation functions into the Laplace domain (Algorithm~\ref{alg:laplace}). Lines 6 and 7 of Algorithm~\ref{alg:laplace} correspond to performing a numerical Laplace transform of a function $f(t)$.
    \item Compute the hydrodynamic friction/thermal resistance,
    \begin{enumerate}
    \item Navier friction coefficient: The Laplace transforms of the correlation functions are passed to the function(s) defined in Algorithm~\ref{alg:fricCoeff}, where the function returns the value of the zero-frequency friction kernel $\zeta_{0}$, which, when divided by the 
    wall surface area, will provide the interfacial friction coefficient ($\xi_{0}$).
    \item Kapitza resistance: The Laplace transform of the correlation functions are passed to the functions, defined in Algorithm~\ref{fun_kap}, where the function returns the value of the Kapitza resistance, $R_k$.
    \end{enumerate}
\end{enumerate}

\begin{algorithm}
\caption{Function algorithm to compute time correlation between two datasets}
\label{alg:corr}
\begin{algorithmic}[1]
\Function{Correlate}{$no\_of\_sets, data1, data2$}
    \State Assign $L \gets \mathrm{LEN}(data1)$
    \State Assign $no\_of\_cols \gets L / no\_of\_sets$
    \State Split $data1$ and $data2$ into $no\_of\_sets$ segments
    \State Define an empty array $corr\_array[\mathrm{ROWS}, \mathrm{COLS}]$,
    \Statex $\mathrm{ROWS} = no\_of\_sets$ and $\mathrm{COLS} = no\_of\_cols$
    \For{each $j$ from 1 to $no\_of\_sets$}
         \State Assign $a \gets data1[j]$, $b \gets data2[j]$
         \State Assign $N \gets \mathrm{LEN}(a)$
         \State Define an empty list $corr$ = []
          \For{each $k$ from 0 to $N-1$}
             \State $Cab \gets \sum_{i=0}^{N-k-1} a[i] \cdot b[i+k] / (N-k)$
             \State Append $Cab$ to $corr$
          \EndFor
         \State Store $corr$ in $corr\_array[\mathrm{ROW}=j,:]$
    \EndFor
    \State Compute $corravg \gets 
    \mathrm{ROW\_MEAN}(corr\_array)$
    \State Compute $corravg\_norm \gets corravg / \mathrm{MAX}(|corravg|)$
    \State \textbf{return} $corravg,\ corravg\_norm$
\EndFunction
\end{algorithmic}
\end{algorithm}

\begin{algorithm}
\caption{Function algorithm to compute numerical Laplace Transform}
\label{alg:laplace}
\begin{algorithmic}[1]
\Function{Laplace}{$time, f\_t, s\_0, s\_{end}, s\_{num}$}
    \State Assign $\Delta t(\mathrm{timestep}) \gets time[1] - time[0]$
    \State Assign $s$ values for the Laplace transform, 
    \Statex $s \gets \mathrm{LINSPACE}(s\_0, s\_{end}, s\_{num})$
    \State Define $Lfs$ with array length = $s\_num$
    \Statex for the Laplace transform data
    \For{each $j$ from 0 to $s\_{num}$}
        \State Assign $e\_{st} \gets exp(-s[j] \cdot time)$
        \State $Lfs[j] \gets \int f\_t \cdot e\_{st}\ \Delta t$ 
    \EndFor
    \State \textbf{return} $s,\ Lfs$
\EndFunction
\end{algorithmic}
\end{algorithm}

\begin{algorithm}
\caption{Function algorithm to compute the Navier friction coefficient}
\label{alg:fricCoeff}
\begin{algorithmic}[1]
\Function{Method1}{$Lcuu, Lcuf$}
    \State Assign $zeta \gets -\, Lcuf/Lcuu$ 
    \State \textbf{return} $zeta[0]$
\EndFunction

\Statex

\Function{Method2}{$frequ, Lcuu, Lcuf$}
    \State Define the fitting model  $func \gets -\dfrac{B}{s + lambda}Lcuu$ 
    \State Perform the regression fitting,
    \Statex $B, lambda \gets \mathrm{CURVE\_FIT}(func,\ (frequ, Lcuu),\ Lcuf)$
    \State Assign $zeta0 \gets B/lambda$
    \State \textbf{return} $zeta0$
\EndFunction

\Statex

\Function{Method3}{$s, Lcuu, Lcuf$}
    \State Assign $s\_mid \gets \mathrm{LEN}(s)/2 $ and $s\_last \gets \mathrm{LEN}(s)$
    \State Assign $\Delta s \gets s[1] - s[0]$
    \State Assign $\Sigma_1 \gets s[0]$;\; $\Sigma_2 \gets s[s\_mid]$;\; $\Sigma_3 \gets s[s\_mid]$;\; $\Sigma_4 \gets s[s\_last-1]$
    \State Assign $A \gets - Lcuu/Lcuf $
    \State Assign $C \gets \dfrac{\int_{0}^{s\_mid} A[s]\ \Delta s}{\int_{s\_mid}^{s\_last} A[s]\ \Delta s}$
    \State $lambda \gets \dfrac{\Sigma_2^2 - \Sigma_1^2 + C(\Sigma_3^2 - \Sigma_4^2)}{2\left(C(\Sigma_4 - \Sigma_3) - \Sigma_2 + \Sigma_1\right)}$
    \State $B \gets \dfrac{\Sigma_2^2 + 2\lambda_1\Sigma_2 - \Sigma_1^2 - 2\lambda_1\Sigma_1}{2\int_{0}^{s\_mid} A[s]\ \Delta s}$
    \State $zeta0 \gets B/lambda$
    \State \textbf{return} $zeta0$
\EndFunction

\end{algorithmic}
\end{algorithm}
\begin{algorithm}
\caption{Function algorithm to compute the Kapitza resistance (proposed EMD framework)}
\label{fun_kap}
\begin{algorithmic}[1]
\Function{Method1}{$LcJqT, LcTT$}
    \State Assign $Gk \gets LcTT/LcJqT$
    \State Assign $Rk \gets 1/Gk$
    \State \textbf{return} $Rk[0]$
\EndFunction

\Statex

\Function{Method2}{$frequ, LcJqT, LcTT$}
    \State Define the fitting model  $func \gets \dfrac{B}{s + lambda}\,LcTT$
    \State Perform the regression fitting,
    \Statex $B, lambda \gets \mathrm{CURVE\_FIT}(func,\ (frequ, LcTT),\ LcJqT)$
    \State Assign $Gk0 \gets B/lambda$
    \State Assign $Rk0 \gets 1/Gk0$
    \State \textbf{return} $Rk0$
\EndFunction

\end{algorithmic}
\end{algorithm}

\newpage
\section{Results and Discussions}
In this section, we present the results computed using the codes described in the previous section. We note that the figures have been reproduced using data from our previous studies on the Navier friction coefficient~\cite{varghese2021} and the Kapitza resistance~\cite{alosious2021}. For the hydrodynamic friction, we consider a planar water–graphene nanochannel, whereas for the thermal friction, we employ a water–CNT system.

\subsection{Hydrodynamic slip}
\begin{figure}[ht]
     \begin{tabular}{cc}
         \includegraphics[width=0.5\linewidth]{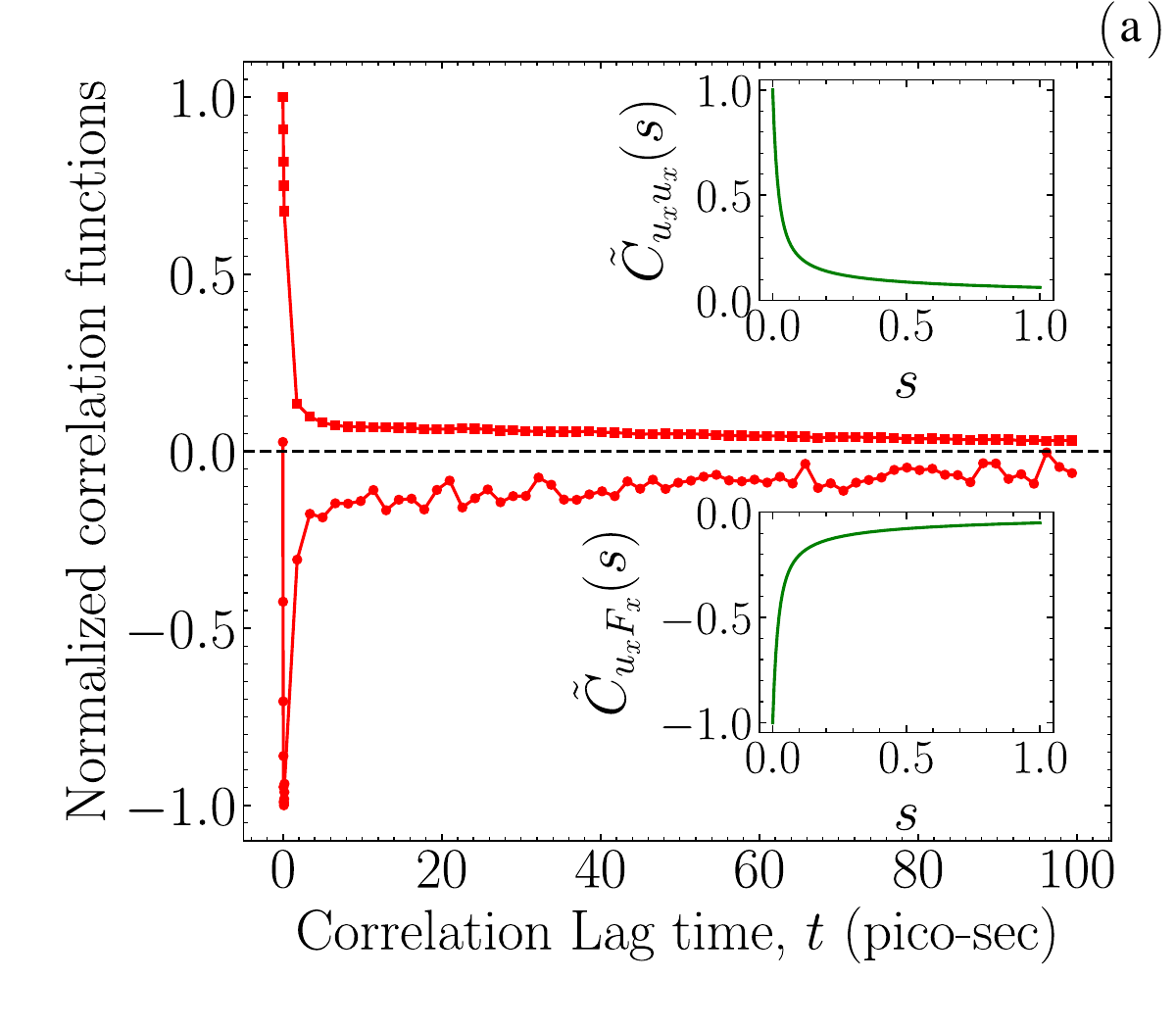} &
         \includegraphics[width=0.46\linewidth]{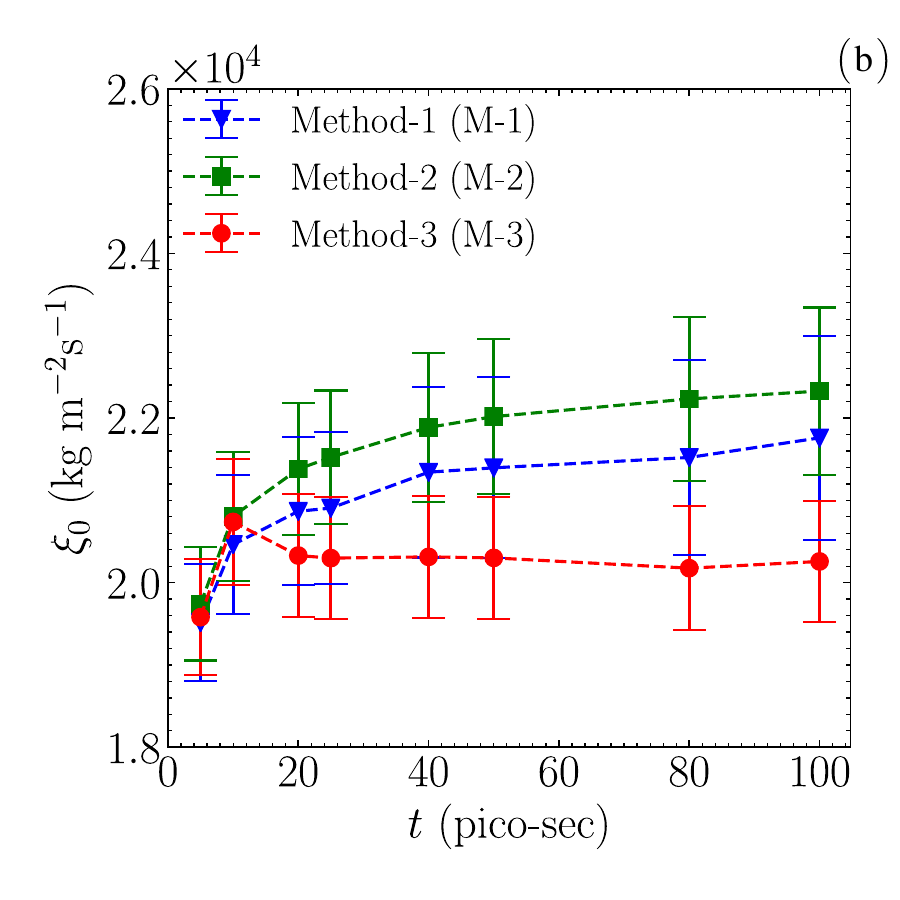} \\
         \includegraphics[width=0.46\linewidth]{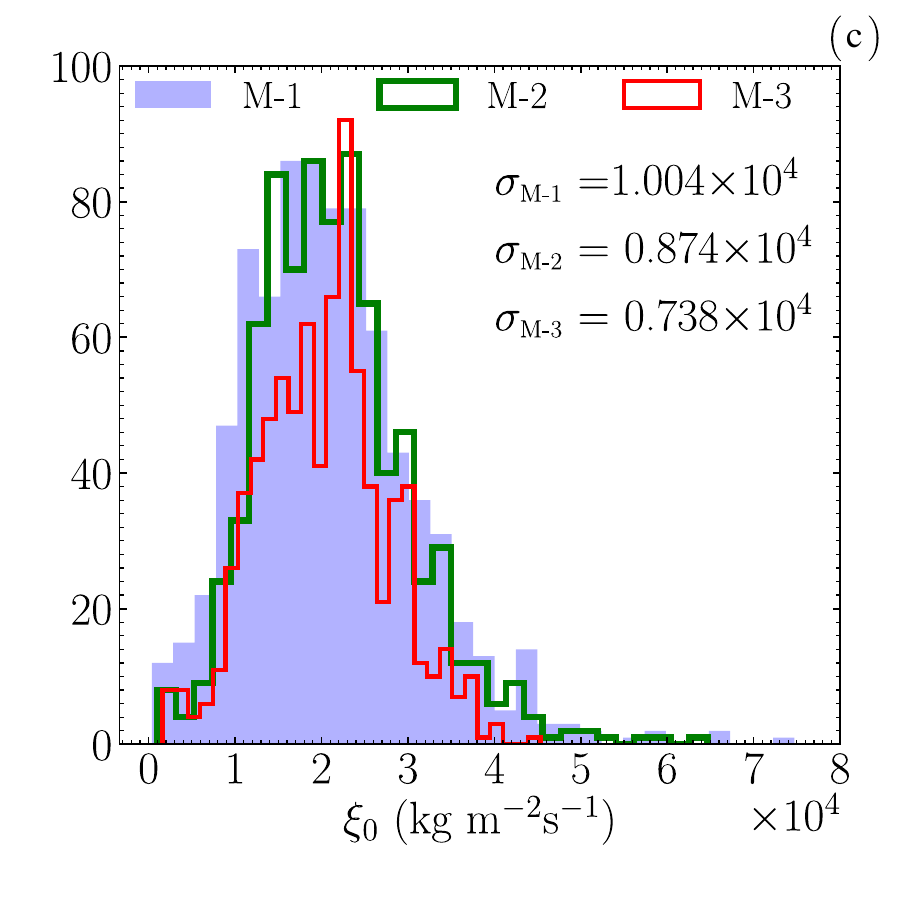} &
         \includegraphics[width=0.46\linewidth]{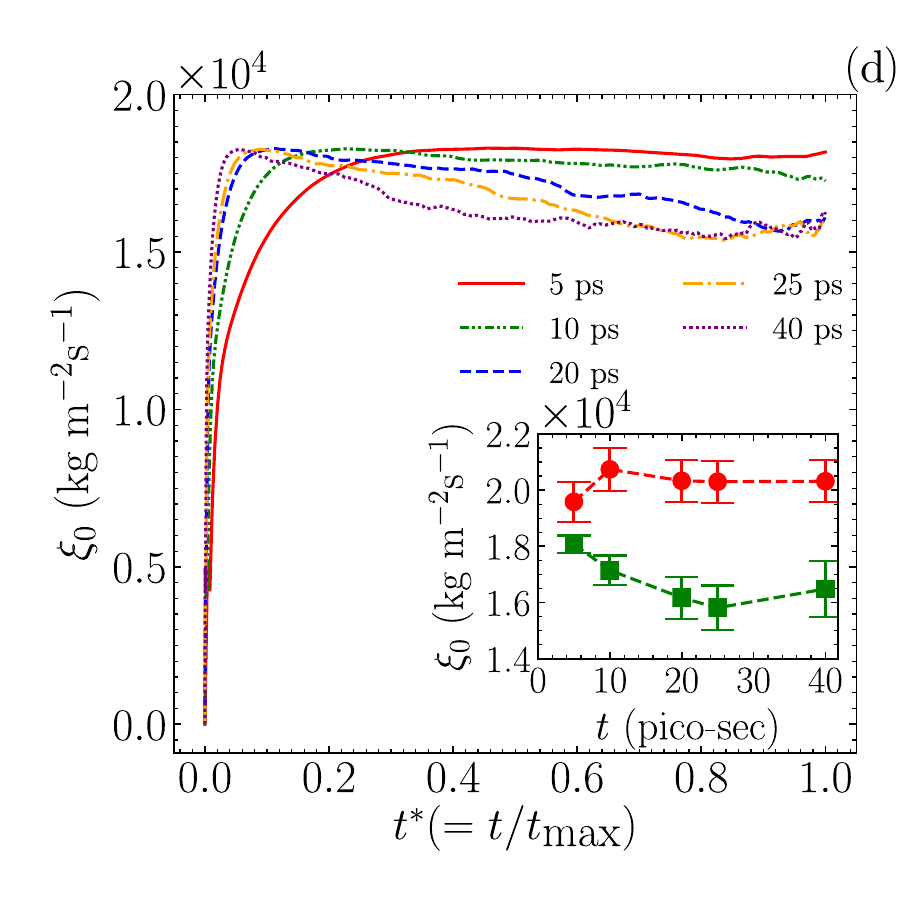}
    \end{tabular}
    \caption{For a water-graphene system: (a) Normalized time correlation functions. Red filled squares represent $C_{u_{x}u_{x}} (t)$, and red filled circles represent $C_{u_{x}F_{x}} (t)$. The inset plot represents the corresponding normalized Laplace transform.
    (b) Variation of friction coefficient with correlation time intervals. The correlation lag times studied are $t$ = 5, 10, 20, 25, 40, 50, 80, and 100 ps. 
    (c) Histogram distribution of friction coefficient values and $\sigma$ represents the standard deviation of the distribution.
    (d) Friction coefficient calculated for different correlation time intervals using the methodology proposed in Ref~\citenum{bocquet1994}. The correlation times in the horizontal axis are normalized with respect to their maximum value for better representation. (Inset plot) Variation of friction coefficient with correlation lag times, for (red filled circle) Method-3 and (green filled square) Ref~\citenum{bocquet1994}. 
    The error bars in (b) and (d) correspond to one standard error obtained from 50 independent simulations.}
    Reprinted with permission from Varghese {\it{et al.}} \cite{varghese2021}. Copyright (2021) American Institute of Physics. 
    \label{fig:hyd_fric_coeff}
     \end{figure}

The first step in the post-processing stage is to compute the time-correlation functions defined in Eq.~\eqref{eq:corr}, and the function algorithm to calculate the time-correlation data using Python is provided in Algorithm~\ref{alg:corr}. Fig~\ref{fig:hyd_fric_coeff} (a) shows an example of the time-correlation functions computed from the time series data of an EMD simulation of a water-graphene system. Further details about the simulation setup are provided in the \ref{app:vel_slip}. Subsequently, the Laplace transforms of the time-correlation data can be computed using the algorithm shown in Algorithm~\ref{alg:laplace}, and inset plots of Fig~\ref{fig:hyd_fric_coeff} (a) show the Laplace transform of the corresponding time-correlation data. 

Though while not specifically mentioned in Sec~\ref{sec:hyd_fric_coeff}, the value for the Navier friction coefficient can be calculated using three different ways. For the first method, referred to as Method-1,
$\zeta_0$ is calculated directly as $\tilde{C}_{u_{x}F_{x}}(0)/\tilde{C}_{u_{x}u_{x}}(0)$, without assuming any functional form for the friction kernel, and the function METHOD1 in Algorithm~\ref{alg:fricCoeff} provides the algorithm to obtain the steady-state value of the friction kernel. For METHOD2 (function METHOD2 in Algorithm~\ref{alg:fricCoeff}) a  
one-term Maxwellian function is assumed for the friction kernel (Eq.~\eqref{friction_kernel}), and the Maxwellian parameters are computed as the fitting coefficients of a regression fit of Eq.~\eqref{reg1}. Finally, Method-3 (METHOD-3 in Algorithm~\ref{alg:fricCoeff}) calculates the parameters in Eq.~\eqref{friction1} from the algebraic expressions given via Eqs.~\eqref{simul1}-\eqref{lambda1}. For Method-3, we chose the integral limits for the Laplacian domain in Eqs.~\eqref{simul1} and \eqref{simul2} as $s_1= 0$, $s_2 = s_3 = 0.5$, and $s_4 = 1$. 
We note that for Method-3, using excessively large intervals in the the Laplacian domain is not ideal, as this effectively incorporates high-frequency statistical noise into the integration. Consequently this can lead to an underestimation of the steady-state friction coefficient. Additionally Method-3 is only applicable to systems whose relaxation dynamics are characterized by a single-exponential friction kernel. For systems exhibiting more complex friction kernels (\emph{e.g.}, confined polymer melts), Method-2 must be used (with appropriate modification of the fitting function), instead of Method-3, to compute the interfacial friction coefficient.

As discussed in the Introduction, one of the commonly cited issues in calculating the Navier friction coefficient from a Green-Kubo-like formalism \cite{bocquet1994} is the issue of the vanishing transport coefficient when integrated over long correlation times. To examine the effect of the correlation lag time in our approach, we compare the friction coefficient values calculated from the correlation functions truncated at different correlation lag times. Fig~\ref{fig:hyd_fric_coeff} (b) plots the friction coefficient values computed at different correlation lag times for a water–graphene system. Although all three methods show similar results for short lag times, the values for $\xi_0$ calculated using Method-1 and Method-2 becomes statistically less reliable as the correlation lag time increases, whereas Method-3 shows consistent values of friction coefficient over the entire range of correlation lag times considered, with smaller statistical errors than Method-1 and Method-2. To quantify the statistical consistency of our methodology, we also plot a histogram distribution of the friction coefficient values over all the correlation lag times in Fig~\ref{fig:hyd_fric_coeff} (c) and observe that the least statistical dispersion is shown by Method-3. This suggests that error propagation in the computation of the friction coefficient is significantly smaller when using Method-3. 
The statistical improvement observed in Method-3 can be attributed to the fact that its variance is governed by the statistical fluctuations of the area under the curve, which are smaller than the fluctuations of the curve itself (Method-2) or those of a data point (Method-1)~\cite{varghese2021}.

We also compare Method-3 with another EMD methodology proposed by Bocquet and Barrat ~\cite{bocquet1994}, which is based on the Green-Kubo formalism, for computing the friction coefficient. An important consideration in the Ref~\citenum{bocquet1994} approach is the choice of the upper limit of the correlation lag time $t$, which should be made judiciously to avoid spurious results for $\xi_0$. This is because the integral of the force autocorrelation function for a finite system decays to zero in the infinite time limit~\cite{espanol1993force, fisher1972friction, smedley1974kirkwood}. A viable workaround to this issue is to consider $\xi_0$ only up to the correlation lag time that shows a plateau for the integrated force autocorrelation function~\cite{falk2010molecular, tocci2014friction}. From Fig.~\ref{fig:hyd_fric_coeff}(d), we can observe that the plateau behavior for the running integral of $\xi_0$ is shown only up to $t = 5$ ps, beyond which there is no observable plateau. Hence, for the Ref~\citenum{bocquet1994} method, we will use the value of friction coefficient at $t = 5$ ps for comparison. In addition, from the inset plots in Fig.~\ref{fig:hyd_fric_coeff}(d), we can see that Method-3 provides more consistent values of friction coefficient across different correlation lag times compared to the Ref~\citenum{bocquet1994} method. 

\subsection{Thermal Slip}
\begin{figure}
    \centering
    \includegraphics[width=0.5\textwidth]{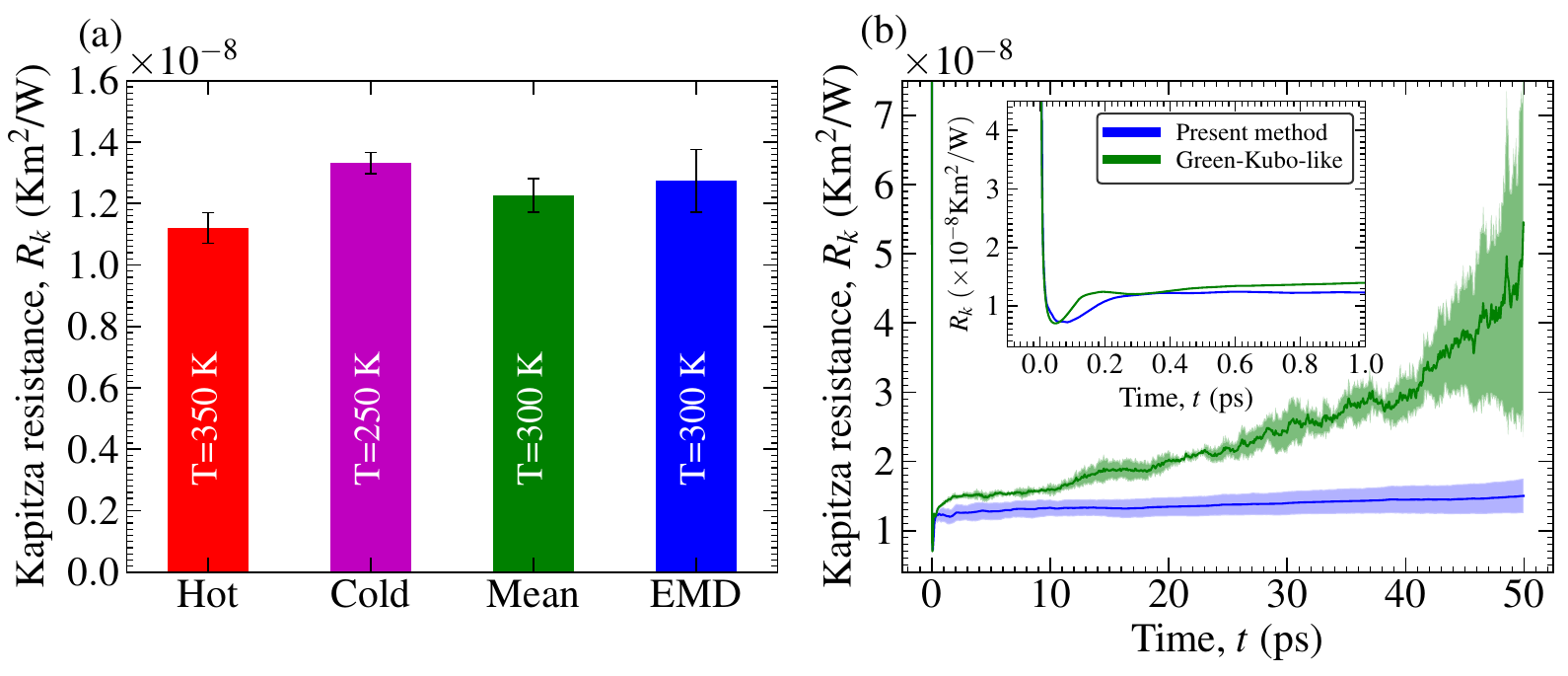}
    \caption{  (a) Comparison of the Kapitza resistance obtained from the NEMD and the
present EMD methods. (b) Comparison of the Kapitza resistance as a function of correlation time calculated using the present method and the Green-Kubo-like method of Barrat and Chiaruttini \cite{puech1986,barrat2003kapitza}. Reprinted with permission from Alosious, {\it{et al.}} \cite{alosious2021}. Copyright (2021) American Chemical Society.
}
    \label{fig:CNT_comparison}
\end{figure}

The calculation of the Kapitza resistance in a cylindrical geometry is demonstrated using a CNT–water system.
 The slab of water adjacent to the CNT wall is defined as an annular region with a thickness of 3.165~\AA, similar to the fluid slab thickness used in the graphene–water system. Figure~\ref{fig:CNT_comparison} (a) compares the Kapitza resistance values obtained from both EMD and NEMD simulations. The EMD method offers a simpler and more flexible approach for cylindrical systems, as it is not limited by the CNT diameter. In contrast, the NEMD method requires additional modeling steps, such as introducing a smaller concentric CNT to generate a temperature gradient, which restricts the range of diameters that can be analyzed. Furthermore, two separate simulations are typically required in NEMD to account for the asymmetric heat flux in cylindrical geometries. Detailed modeling and simulation procedures for the NEMD calculations are available in our previous work~\cite{alosious2021}. The excellent agreement between the EMD and NEMD results confirms the validity and robustness of the present methodology for computing Kapitza resistance in cylindrical geometries.
 
 Barrat and Chiaruttini developed an EMD approach based on the Green–Kubo formalism to evaluate Kapitza resistance for planar interfaces. In this work, we benchmark our EMD approach against their formulation and observe that our method does not suffer from the well-known plateau problem associated with Green–Kubo formalism of confined fluids \cite{espanol2019solution}. Figure~\ref{fig:CNT_comparison} (b) presents a comparison of the Kapitza resistance as a function of correlation time obtained using our EMD approach and the Green–Kubo-type method introduced by Barrat and Chiaruttini \cite{puech1986,barrat2003kapitza}. While both approaches yield consistent values at short correlation times, the Green–Kubo-based method exhibits a divergence as the correlation time increases, whereas the results from our EMD method remain largely insensitive to the choice of correlation time. This robustness arises because our formulation directly correlates interfacial heat flux with temperature-difference fluctuations, thereby leveraging a larger fraction of the available simulation data. Furthermore, since the correlation functions are evaluated within an interfacial fluid region of thickness $\Delta = 3.165~\text{\AA}$ (measured from the CNT surface to the first peak in the water density profile), the resulting Kapitza resistance represents a local interfacial property rather than a bulk fluid property.

The finite Kapitza resistance arises from the mismatch between the phonon modes of the CNT wall and the vibrational modes of the confined water molecules, which limits the efficiency of interfacial energy transfer. Heat transport across the interface is therefore governed by the coupling between CNT phonon modes and the vibrational modes of the confined water molecules within the first interfacial water layer. The ability of the present methodology to isolate this local interfacial contribution is particularly important in nanoscale systems, where interfacial thermal resistance can dominate the overall heat-transfer process.

\section{Conclusions}
We have presented a computational framework from which the Navier friction coefficient (for velocity slip) and the Kapitza resistance (for thermal slip) can be evaluated using molecular dynamics simulations. The method is based on calculating the fluid equilibrium fluctuations in slabs adjacent to the walls, i.e., the local properties of the wall-fluid system are extracted. This, therefore, overcomes the size dependent problem present in other global system approaches. We have provided details showing how the relevant data can be extracted using LAMMPS. We also give the reader detailed algorithms for post-processing of the data to compute the Navier friction coefficient and Kapitza resistance.

The primary strengths of the NAVIS framework lie in its ability to compute intrinsic interfacial transport coefficients directly from equilibrium molecular dynamics (EMD) simulations and its applicability to both planar and cylindrical geometries. The method demonstrates superior accuracy and flexibility compared to existing EMD approaches, particularly in calculating Kapitza resistance in cylindrical geometries without the need for an auxiliary cylinder. Additionally, the present EMD method avoids the limitations of standard NEMD approaches for slip computation, which require extrapolation to zero strain rate or thermal gradient -- a procedure that is particularly challenging for high-slip systems.

The main limitation of the method stems from its reliance on linear-response theory, i.e., the transport properties obtained using this method can only be meaningfully compared with experimental measurements performed under conditions where the applied shear rates or thermal gradients remain within the linear regime. Another limitation is that the primary outputs of the methodology are the interfacial friction coefficient and the Kapitza resistance. Experimental studies, however, often report the corresponding slip length and Kapitza length. Reliable conversion of the computed interfacial friction coefficient and Kapitza resistance into slip and Kapitza lengths requires accurate knowledge of the fluid viscosity and thermal conductivity at the solid–liquid interface, quantities that may not always be readily available.

Nevertheless, the framework provides direct access to the fundamental interfacial transport coefficients that govern momentum and heat transfer at solid-liquid interfaces.
Since accurate boundary conditions are critical for modeling nanoscale transport phenomena ~\cite{hansen:book:2022}, we expect the NAVIS framework to offer researchers an efficient and robust tool for determining these quantities.

\appendix
\section{Simulation details}
\subsection{Navier friction coefficient}
\label{app:vel_slip}
\begin{figure}
    \centering
    \includegraphics[width=1.0\linewidth]{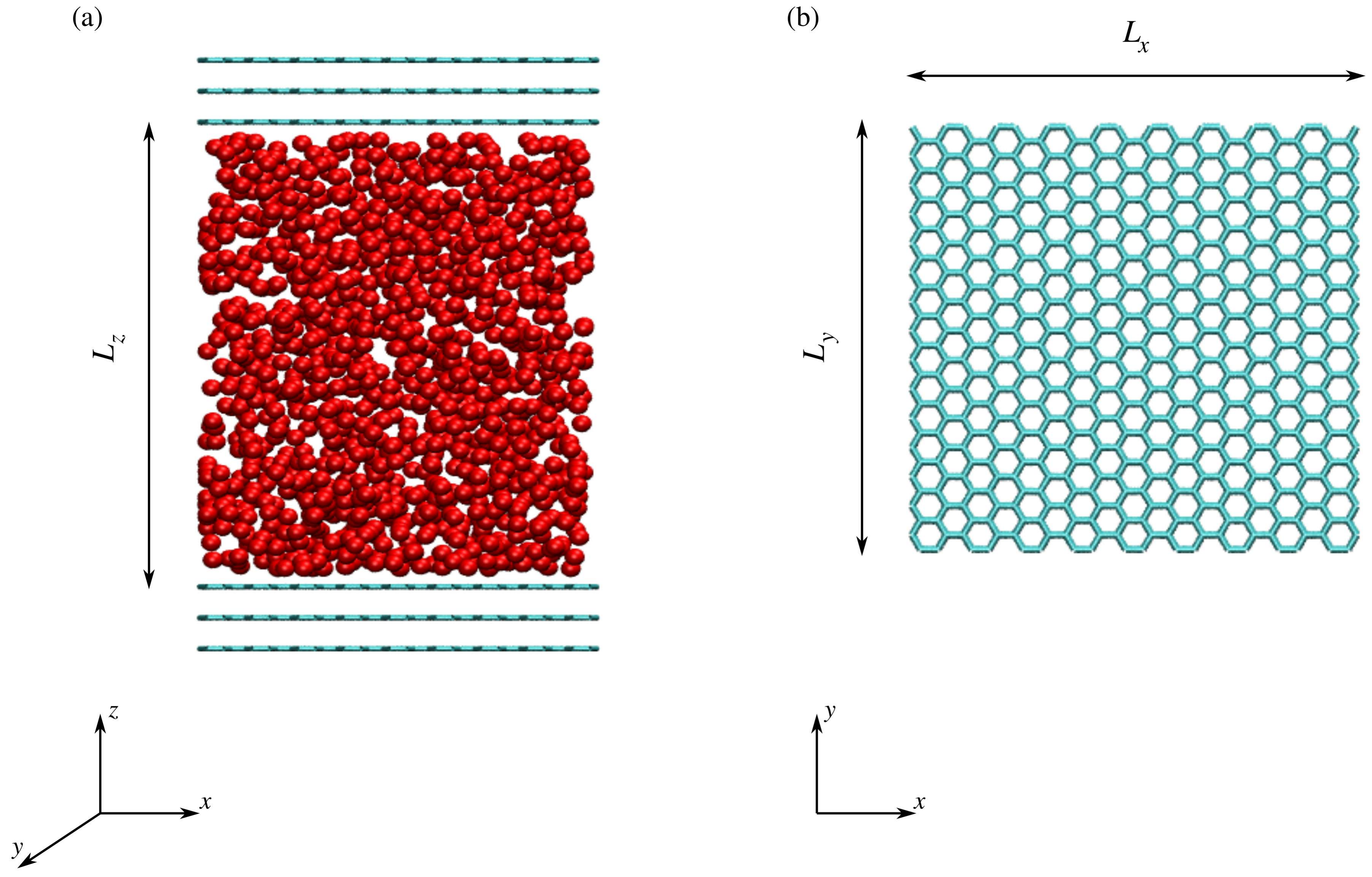}
    \caption{(a) Schematic representation of the planar confined system. (b) Top view of the graphene channel. Reprinted with permission from Varghese {\it{et al.}} \cite{varghese2021}. Copyright (2021) American Institute of Physics.}
    \label{fig:wat_gra_model}
\end{figure}

The schematic illustration of the water-graphene system used for the EMD simulations is shown in Fig~\ref{fig:wat_gra_model}. In  Fig~\ref{fig:wat_gra_model} the graphene sheets are placed in the $x-y$ plane with periodicity in both directions and confinement along the $z$ axis. Each wall consists of three layers of stacked uncharged graphene sheets kept at an interlayer distance of $3.4\ \AA$. The outermost layer at each wall is kept fixed to maintain a constant system size. The dimensions for the graphene nanochannel in the $x-y$ plane are $L_{x} = 36.87\ \AA$ and $L_y = 35.61\ \AA$. The intralayer carbon interactions for the graphene sheets were modeled using the optimized Tersoff potential~\cite{tersoff1989modeling, lindsay2010optimized}. A weak Lennard-Jones potential is also applied between the interlayer carbon atoms to hold the graphene sheets together~\cite{kannam2011, kannam2012}.

To model the water molecules we use SPC/E~\cite{berendsen1987missing, wu2006flexible} water model with the bond and angles of each water molecule constrained using the SHAKE~\cite{ryckaert1977numerical} algorithm in LAMMPS. Each hydrogen atom in the water molecule carries a partial charge of $0.4238e$, and an oxygen atom carries a partial charge of $-0.8476e$, where $e$ is the elementary charge of an electron. 
The van der Waals parameters for water-carbon (WC) interaction ($\epsilon_{\textrm{WC}}$ = 0.09369 kcal/mol, $\sigma_{\textrm{WC}}$ = 3.19 \AA) are taken from Werder et al~\cite{werder2003water}. The Lennard-Jones  parameters for oxygen-oxygen (OO) interaction are $\epsilon_{\textrm{OO}} = 0.15535$  kcal/mol and $\sigma_{\textrm{OO}}$ = 3.166 \AA, and Lennard-Jones interaction  coefficients are zero for hydrogen atoms. The interaction cutoffs for Lennard-Jones and short-range Coulombic interactions were both kept at 10 \AA. The long-range electrostatic interactions are calculated using the Ewald algorithm with the particle-particle-particle-mesh solver~\cite{hockney2021computer} of LAMMPS with a relative root mean square error in the per-atom force calculations below $10^{-6}$. To accommodate the non-periodicity in the $z$ direction, we use the corrected Ewald algorithm EW3DC~\cite{yeh1999ewald}, where the ratio of extended volume to actual channel size is set as 3.0.

The equations of motion for all particles are integrated using the Velocity-Verlet~\cite{swope1982computer} scheme of LAMMPS with an integration time step $\Delta t$ = 1 fs. To maintain an internal pressure of 1 bar the channel width is adjusted by fixing the bottom wall and using the top wall as a piston to apply pressure at 300 K. After setting the internal pressure, the water–graphene system is simulated for a total runtime of 6 ns. During the initial 2.0 ns, the system is equilibrated at 300 K by thermostating the walls using the Langevin thermostat~\cite{allen2017computer}. The equilibration stage is followed by a 2.0 ns run in the NVE ensemble to verify the system stability, and the data required for the post-processing stage are collected from the final 2.0 ns run.

\subsection{Slab width}
The slab width ($\Delta$) was selected to capture the first interfacial fluid layer adjacent to the solid surface. For the SPC/E water systems considered here, $\Delta\ (= 3.165$~\AA)\ is approximately equal to the Lennard--Jones diameter of the water oxygen atom, and corresponds closely to the thickness of the first adsorbed water layer near the wall. More generally, $\Delta$ should be regarded as a system-dependent parameter and selected from the equilibrium density profile normal to the surface. We recommend choosing $\Delta$ such that the slab encompasses the first interfacial adsorption layer, typically extending to the first minimum following the first density peak. The robustness of the selected value can be verified through a sensitivity analysis to ensure that the computed transport coefficients remain statistically invariant to small changes in slab width~\cite{hansen2011}.

\subsection{NEMD simulations (Friction coefficient)} 
\begin{figure}
    \centering
    \includegraphics[width=0.7\linewidth]{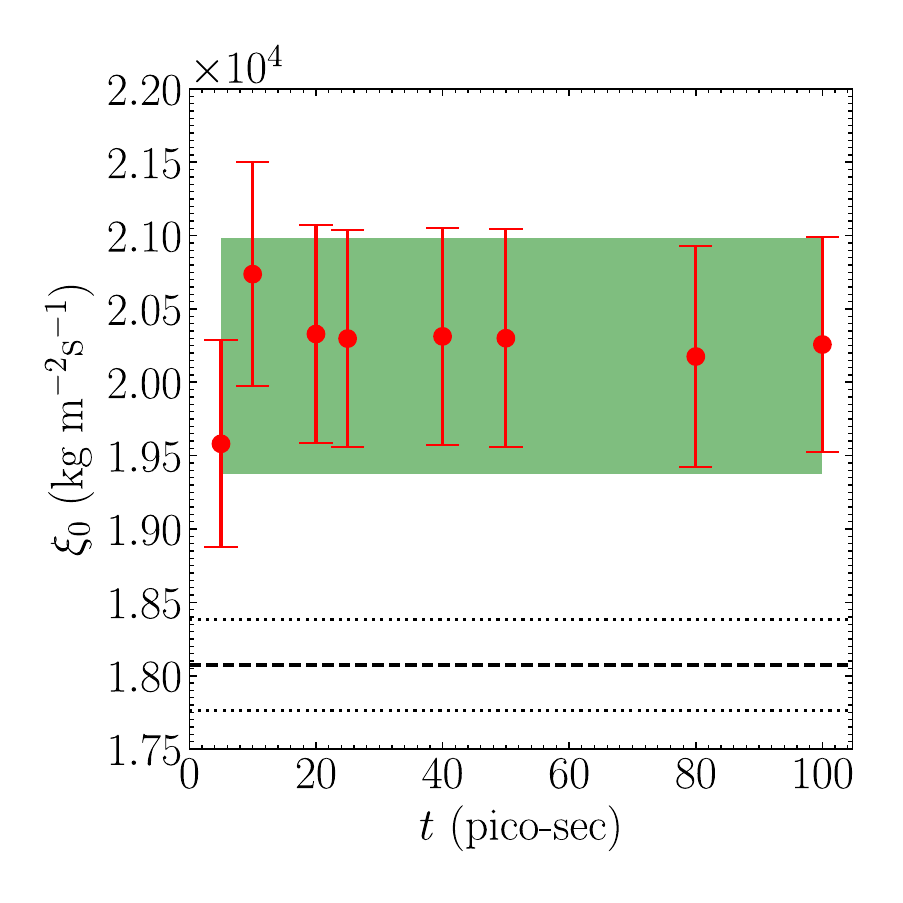}
    \caption{
    (e) Comparison of $\xi_0$ calculated using different methods. Red filled circles represent $\xi_0$ calculated using Method-3 at different correlation lag intervals. The dashed line shows the $\xi_0$ calculated using the methodology proposed in Ref \citenum{bocquet1994} at the correlation lag time $t$ = 5 ps, and the dotted lines correspond to the upper and lower limits of the friction coefficient. The green shaded region represents the NEMD friction coefficient range at the external field $\sim 1.25 \times 10^{11}\ \textrm{m}\textrm{s}^{-2}$.
    Reprinted with permission from Varghese {\it{et al.}} \cite{varghese2021}. Copyright (2021) American Institute of Physics.}
    \label{fig:nemd}
\end{figure}

We note that in our original work \cite{varghese2021} we also compared the EMD results with the friction coefficient values computed from NEMD simulations. For the NEMD calculations, we simulated a Hagen–Poiseuille flow by applying an external field to the confined fluid and Fig~\ref{fig:nemd} shows the comparison of the friction coefficient values using different methodologies for a water-graphene system. From Fig~\ref{fig:nemd} we can see that Method-3 reliably predicts the friction coefficient estimated from NEMD simulations, thereby confirming the validity of our EMD methodology.

\subsection{Kapitza Resistance}

\begin{figure}[h]
	\centering
	\includegraphics[width=0.7\linewidth]{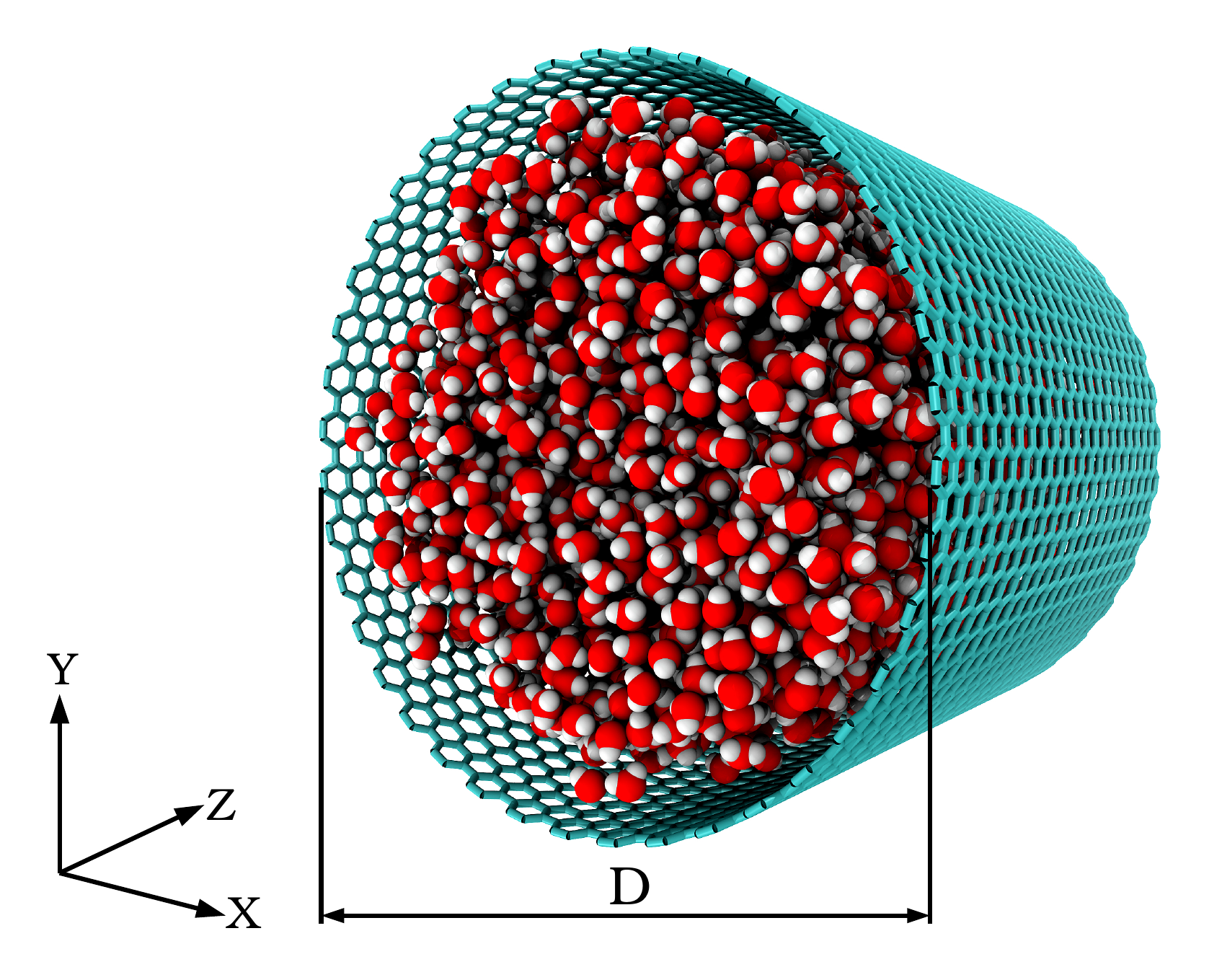}
	\caption{ Schematic diagram of water-CNT system. Reprinted with permission from Alosious, {\it{et al.}} \cite{alosious2021}. Copyright (2021) American Chemical Society.}
	\label{cnt}
\end{figure}
The cylindrical system consists of water molecules confined inside a CNT and the schematic depiction of the initial configuration is shown in Fig. \ref{cnt}. A (30,30) chiral CNT with a diameter of 4.06 nm is used in this particular case. The diameter of a ($n$,$m$) CNT can be calculated using the equation $d=\sqrt{3}l/\pi\sqrt{n^2+nm+m^2}$ where $l$=0.142 nm is the carbon-carbon bond length. The initial configuration of the water-CNT system is created by immersing a rigid CNT in a water bath and equilibrating at 300K temperature and 1 bar pressure. This process will ensure the proper filling of water molecules inside the CNT. The final equilibrated system after removing the water molecules outside of CNT is used as the initial configuration for MD simulation. 
The water model used here is the same as that employed in the water–graphene system for the friction coefficient. The equations of motion for water and CNT were integrated using a leap-frog integration scheme with a time step $\Delta t$ = 1 fs. The simulation process involves minimization of the system followed by equilibration in canonical (NVT) ensemble for 2 ns at 300K temperature. Following that, the system stability was verified by a microcanonical (NVE) ensemble simulation for another 2 ns. For the production run, a thermostat is maintained at 300 K on the CNT for 5 ns, during which the necessary data is extracted.  

\section{Comparison with existing methods}
\begin{table}[ht]
    \centering
    \begin{tabular}{l|c|c|c}
         Observable &  NAVIS & G-K & NEMD \\
         \hline 
         Friction coefficient  & & & \\
         ($\times 10^4\ \mathrm{kg}\ \mathrm{m}^{-2}\ \mathrm{s}^{-1}$) & $(2.03 \pm 0.07)$ & $(1.81 \pm 0.03)$ & $(2.01 \pm 0.08)$ \\
         \hline 
         Kapitza resistance & & & \\
         ($\times 10^{-8}\ \mathrm{K}\ \mathrm{m}^{2}\ \mathrm{W}^{-1}$) & $(1.27 \pm 0.09)$ & $(1.28 \pm 0.03)$ & $(1.24 \pm 0.02)$  
    \end{tabular}
    \caption{Table comparing the friction coefficient and Kapitza resistance values obtained using different methods. For the friction coefficient, the NAVIS value corresponds to $t = 40$ ps and G-K value corresponds to $t = 5$ ps.}
\end{table}

\section{GitHub folder structure}
\label{sec:App_rep}
\begin{forest}
  for tree={
    font=\ttfamily,
    grow'=0,
    child anchor=west,
    parent anchor=south,
    anchor=west,
    calign=first,
    edge path={
      \noexpand\path [draw, \forestoption{edge}]
      (!u.south west) +(7.5pt,0) |- node[fill,inner sep=1.25pt] {} (.child anchor)\forestoption{edge label};
    },
    before typesetting nodes={
      if n=1
        {insert before={[,phantom]}}
        {}
    },
    fit=band,
    before computing xy={l=15pt},
  }
[NAVIS
  [Navier-friction-coefficient
  [LAMMPS
    ]
    [POST-PROCESSING
        [customPkgs]
        [bot
        [correlation]
        [friction-coefficient]
        ]
        [top
        [correlation]
        [friction-coefficient]
        ]
        ]
  ]
  [Kapitza-resistance
    [LAMMPS
    ]
    [POST-PROCESSING]
  ]
]
\end{forest}




\section*{Acknowledgments}
The results presented in this work was performed on the OzSTAR national facility at Swinburne University of Technology. The OzSTAR program receives funding in part from the Astronomy National Collaborative Research Infrastructure Strategy (NCRIS) allocation provided by the Australian Government, and from the Victorian Higher Education State Investment Fund (VHESIF) provided by the Victorian Government.

\section*{Author declarations}
\subsection*{Conflict of interest}
There is no conflict of interest to declare.


\bibliographystyle{elsarticle-num}
\bibliography{references}







\end{document}